# Intrinsic Radial Landau Rainbow in Triaxial Strained Photonic Crystal

**GUANGTAI LU,[1,] * SATOSHI IWAMOTO,[1, 2]**

[1] *Research Center for Advanced Science and Technology, The University of Tokyo, 4-6-1 Komaba, Meguro-ku, Tokyo 153-8904, Japan*

[2] *Institute of Industrial Science, The University of Tokyo, 4-6-1 Komaba, Meguro-ku, Tokyo 153-8904, Japan*

**lugt@iis.u-tokyo.ac.jp*

## Abstract

Localized resonances are essential for enhancing light–matter interaction, controlling emission, and realizing compact optical resonators and lasers. Landau levels offer a distinct route to organizing photonic states through synthetic magnetic fields, but their high degeneracy does not define spectrally separated localized resonances. Here, we report the intrinsic lifting of degeneracy in photonic Landau levels in a triaxially strained photonic crystal, leading to the emergence of an intrinsic radial Landau rainbow. The triaxial deformation generates photonic Landau levels through a strain induced pseudomagnetic field, while an accompanying pseudoelectric field lifts the degeneracy of the Landau levels. As a result, each Landau level splits into an equally spaced frequency ladder of localized resonances. In zeroth Landau level, these modes exhibit intensity maxima that move progressively outward from the device center, establishing a radial frequency–position mapping and Landau rainbow. Our results reveal an intrinsic fine structure of photonic Landau levels in a triaxial strained photonic crystal and provide a route to spatially ordered Landau level resonances.

## Introduction

Synthetic gauge fields provide a powerful route to control classical waves by reproducing magnetic-field-like responses in engineered media.[1–3] A particularly important example is the strain induced pseudomagnetic field in honeycomb lattices, where nonuniform deformation shifts the Dirac points in momentum space and reorganizes the Dirac spectrum into Landau levels.[4,5] Originating from strain engineering in graphene[6–9], this concept has been extended to a broad range of classical wave platforms, including photonic[10–12], acoustic[13,14], and polaritonic[15] systems. Among various deformation configurations, triaxial strain is widely studied in generating an approximately uniform pseudomagnetic field in a centrally organized geometry. In photonic crystals, triaxially deformed honeycomb lattices have recently been used to realize Landau-quantized optical resonances and their electrical tuning[16].

However, Landau quantization alone does not in general define localized resonances. This issue is important in photonics, where spatially confined resonances are central to enhancing light–matter interaction.[17–19] In an ideal uniform magnetic or pseudomagnetic field, each Landau level retains a degeneracy: many states share the same eigenfrequency while being associated with different spatial distributions. Additional physical effects are therefore required to resolve this degeneracy into modes that are both spectrally distinguishable and spatially organized.

Recently, two distinct mechanisms for unfolding photonic Landau level degeneracy have been discussed. First, Landau rainbow systems have demonstrated that an engineered pseudoelectric field associated with a scalar potential can lift the degeneracy of Landau modes and map different frequencies to different spatial positions.[20,21] In the reported configurations, this frequency–position mapping is predominantly directional, producing a lateral displacement of the Landau modes along an imposed gradient. Second, lateral strained photonic crystals have been shown to support dispersive Landau levels, in contrast to the flat levels predicted by an ideal tight-binding description.[22–24] This dispersion has been attributed to higher-order corrections inherent to the continuum system.[22] In geometries retaining a translationally invariant direction, such corrections produce Landau modes that are confined in the transverse direction while remaining labelled by the propagation wavevector. These developments suggest that, in a centrally organized triaxially strained photonic crystal, the corresponding higher-order correction may instead unfold a Landau level into a discrete set of radially localized resonances.

Here, we report an intrinsic radial Landau rainbow in a triaxially strained graphene-like photonic crystal. We show that the higher-order correction identified in previous studies can be described by an accompanying pseudoelectric field. While the strain induced pseudomagnetic field establishes the Landau level spectrum, this accompanying pseudoelectric field lifts the degeneracy of Landau levels and resolves them into a sequence of discrete resonant modes. These modes form an approximately equally spaced spectral ladder and exhibit intensity maxima that move progressively outward from the device center, providing a radial rainbow configuration. Our results identify the accompanying pseudoelectric field as the origin of the intrinsic fine structure of photonic Landau levels and establish a radial form of Landau rainbow without requiring an externally imposed separation mechanism.

## Method

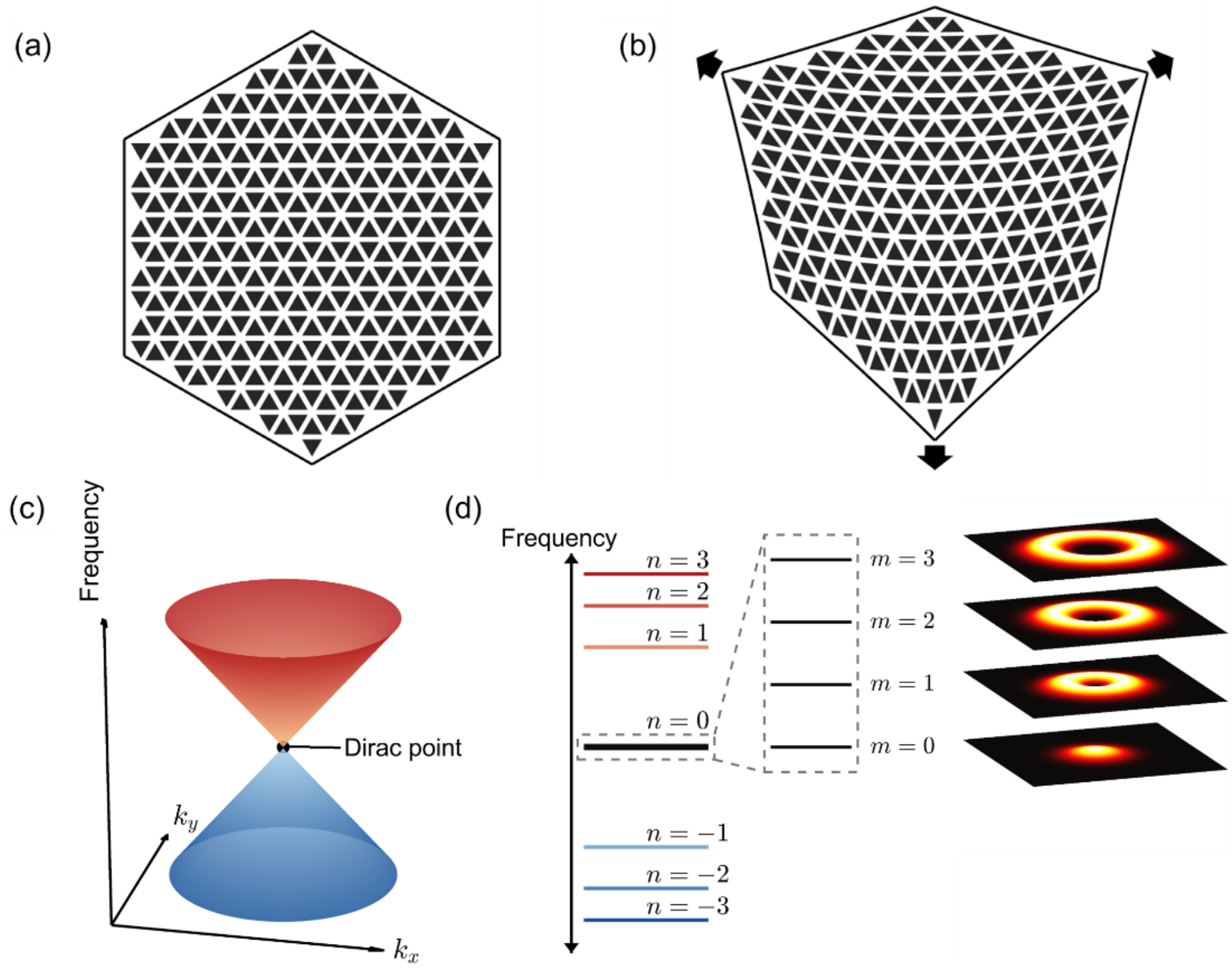

**Figure 1. Concept of the intrinsic radial Landau rainbow in a triaxially strained graphene-like photonic crystal.** (a) Unstrained graphene-like photonic crystal with a finite hexagonal geometry. (b) Triaxially strained photonic crystal obtained by deforming the lattice along three directions. (c) Dirac cone of the unstrained photonic crystal in momentum space. (d) Schematic eigenfrequency spectrum of the triaxially strained photonic crystal. The Dirac spectrum quantizes into Landau levels, with a sequence of spectrally separated sublevels. These split modes have different radial intensity maxima, forming an intrinsic radial Landau rainbow.

We consider a two-dimensional graphene-like photonic crystal consisting of triangular air holes arranged in a honeycomb lattice (Figure 1a). The lattice constant is $a = 500\,\mathrm{nm}$, and the side length of each triangular hole is $300\,\mathrm{nm}$. The slab is described by a two-dimensional effective index approximation, with $n_{\mathrm{eff}} = 2.6$ for the dielectric background and $n = 1$ for the air holes.

To realize a finite strained structure, we construct a hexagonal photonic crystal region with 40 lattice periods along each side. Synthetic triaxial strain is illustrated in Figure 1b. The strain can be described by the following spatial shift in the structure

$$\boldsymbol{r}' = \boldsymbol{r} + \boldsymbol{u}(\boldsymbol{r}) \qquad (1)$$

where the displacement field is

$$u_x = 2\frac{C}{a}xy$$
$$u_y = \frac{C}{a}(x^2 - y^2) \quad (2)$$

Here the coordinate origin is chosen at the center of the hexagonal region, and $C$ is a dimensionless parameter controlling the strain strength. This deformation is the photonic analogue of the triaxial strain field used to generate a uniform pseudomagnetic field in strained graphene.

Within the effective Dirac description, the strain induced vector potential at the $K$ valley can be written as

$$A_x = \frac{\beta}{a}\left(\partial_x u_x - \partial_y u_y\right) = \frac{4\beta C y}{a^2}$$
$$A_y = -2\frac{\beta}{a}\partial_y u_x = -\frac{4\beta C x}{a^2} \quad (3)$$

where $\beta$ is an effective strain-coupling coefficient. Therefore, the corresponding pseudo-magnetic field is spatially uniform:

$$B_{ps} = \partial_x A_y - \partial_y A_x = -\frac{8\beta C}{a^2} \quad (4)$$

The sign of $B_{ps}$ is reversed at the $K'$ valley, as required by time-reversal symmetry.

Near the Dirac frequency, the effective Hamiltonian can be written as

$$H = v_D\left(\sigma_x \Pi_x + \sigma_y \Pi_y\right)$$
$$\Pi_i = -j\partial_i + A_i \quad (5)$$

where $v_D$ is the Dirac velocity, $A_i$ is the strain-induced vector potential, and its eigenvalue $\omega$ denotes the frequency detuning from the Dirac point. For a uniform pseudomagnetic field, it is well known that the original Dirac cone (Figure 1c) is quantized into a series of Landau levels (Figure 1d),

$$\omega_n = sgn(n)\frac{v_D}{l_B}\sqrt{2|n|}, \qquad n = 0, \pm 1, \pm 2, \ldots \quad (6)$$

where $l_B$ is the magnetic length defined as

$$l_B = 1/\sqrt{|B_{ps}|} \quad (7)$$

A more detailed discussion is provided in Supplementary Information S1 and S2.

In photonic crystals, however, the strain induced vector potential is accompanied by a scalar potential term arising from the deformation of the local unit cell geometry, which shifts the Dirac frequency. In the triaxially strained photonic crystal, the first-order correction of Dirac frequency vanishes because the strain tensor is traceless.[4] Therefore, the leading rotationally symmetric contribution to the scalar potential arises from the second-order strain invariant, which gives a quadratic radial form near the center of the device (see Supplementary Information S3):

$$V(r) = \kappa r^2 \quad (8)$$

where $\kappa$ is a constant for a given strain strength $C$. This scalar potential, which corresponds to a pseudoelectric field, lifts the degeneracy of each Landau level, and gives a frequency splitting. For zeroth Landau level, this splitting is linear,

$$\Delta\omega_{0,m} \sim 2\kappa l_B^2 (m + 1/2), \qquad m = 0,1,2,\ldots \quad (9)$$

where $m$ is a quantum number and is related to the angular momentum of the state. A detailed discussion is provided in Supplementary Information S3. The resulting energy levels and field distribution are shown in Figure 1d. States with larger $m$ have larger radial extent, their intensity maxima move outward, giving rise to the radial Landau rainbow structure discussed below.

## Results

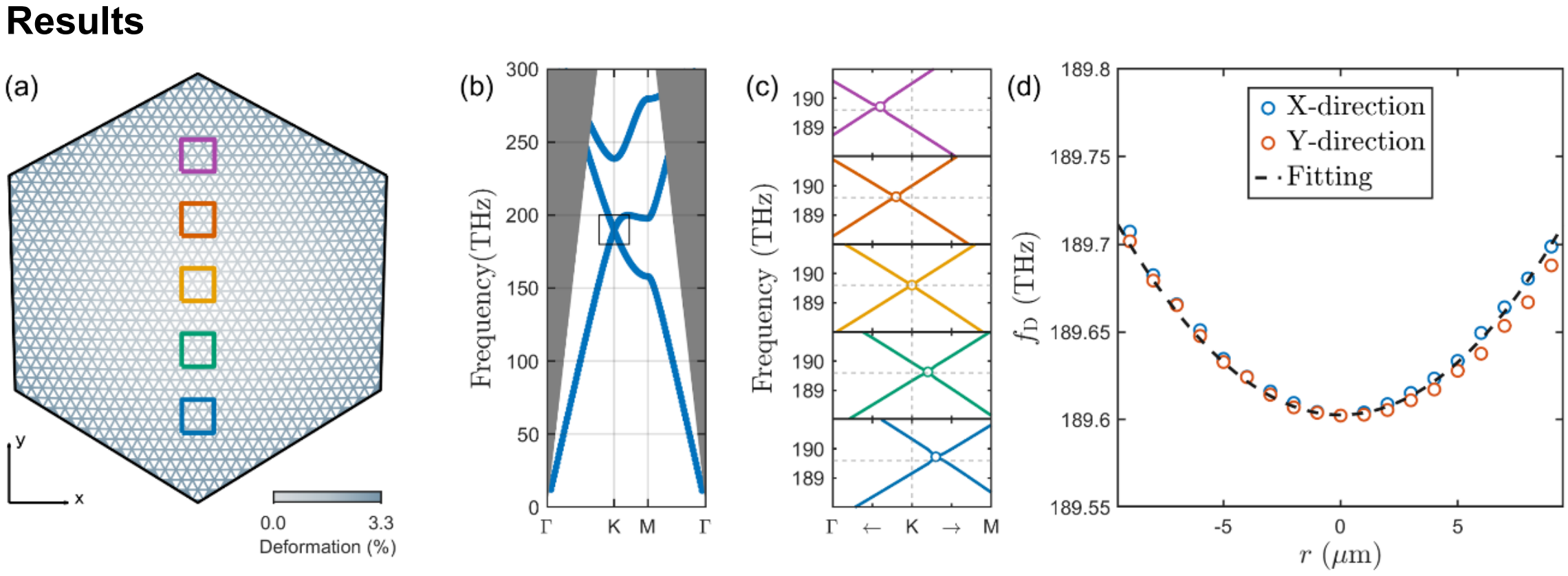

**Figure 2. Local band structure in the triaxially strained photonic crystal.** (a) Triaxial strained photonic crystal with strain strength $C = 5 \times 10^{-4}$. The background color represents the local deformation strength, and the five colored squares mark the positions used for the local band structure calculations in (c). (b) Band structure of the unstrained photonic crystal, showing the Dirac cone at K point. (c) Local band structures near K point for the positions indicated in (a). (d) Extracted local Dirac frequency as a function of radial position along the X- and Y-directions. The quadratic fit indicates a radial scalar potential accompanying the triaxial deformation.

We numerically analyze the strained photonic crystals using COMSOL Multiphysics within the two-dimensional effective index model, as illustrated in Figure 2a. We first calculate the band structure of the unstrained structure. The results, shown in Figure 2b, exhibit a gapless Dirac cone at the valleys around 190 THz, providing a photonic analogue of a honeycomb Dirac system. The Dirac velocity is extracted from the linear dispersion by the relationship $v_D = \partial\omega / \partial k$. The extracted value is $v_D = 8.8 \times 10^7$ m/s.

Although the globally strained structure is aperiodic, the deformation varies slowly on the scale of the lattice constant, allowing us to define a locally periodic approximation at each position. Figure 2c shows representative local band structures near the Dirac point for unit cells extracted at different radial positions. The local deformation produces two distinct modifications of the Dirac cone. First, the calculated Dirac point exhibits a position-dependent displacement relative to the original Brillouin zone corner. This relative momentum shift represents the effective vector potential responsible for the pseudomagnetic field. Second, the frequency of the local Dirac frequency shifts systematically with position. Unlike the ideal tight-binding graphene model, the Dirac frequency is sensitive to deformation induced changes in its local unit cell geometry, giving rise to a scalar contribution to the effective Dirac description. As shown in Figure 2d, the extracted Dirac frequency shift follows a quadratic radial dependence. This behavior identifies an effective radial pseudoelectric field. The local band analysis therefore provides microscopic support for the mechanism of accompanying pseudoelectric field.

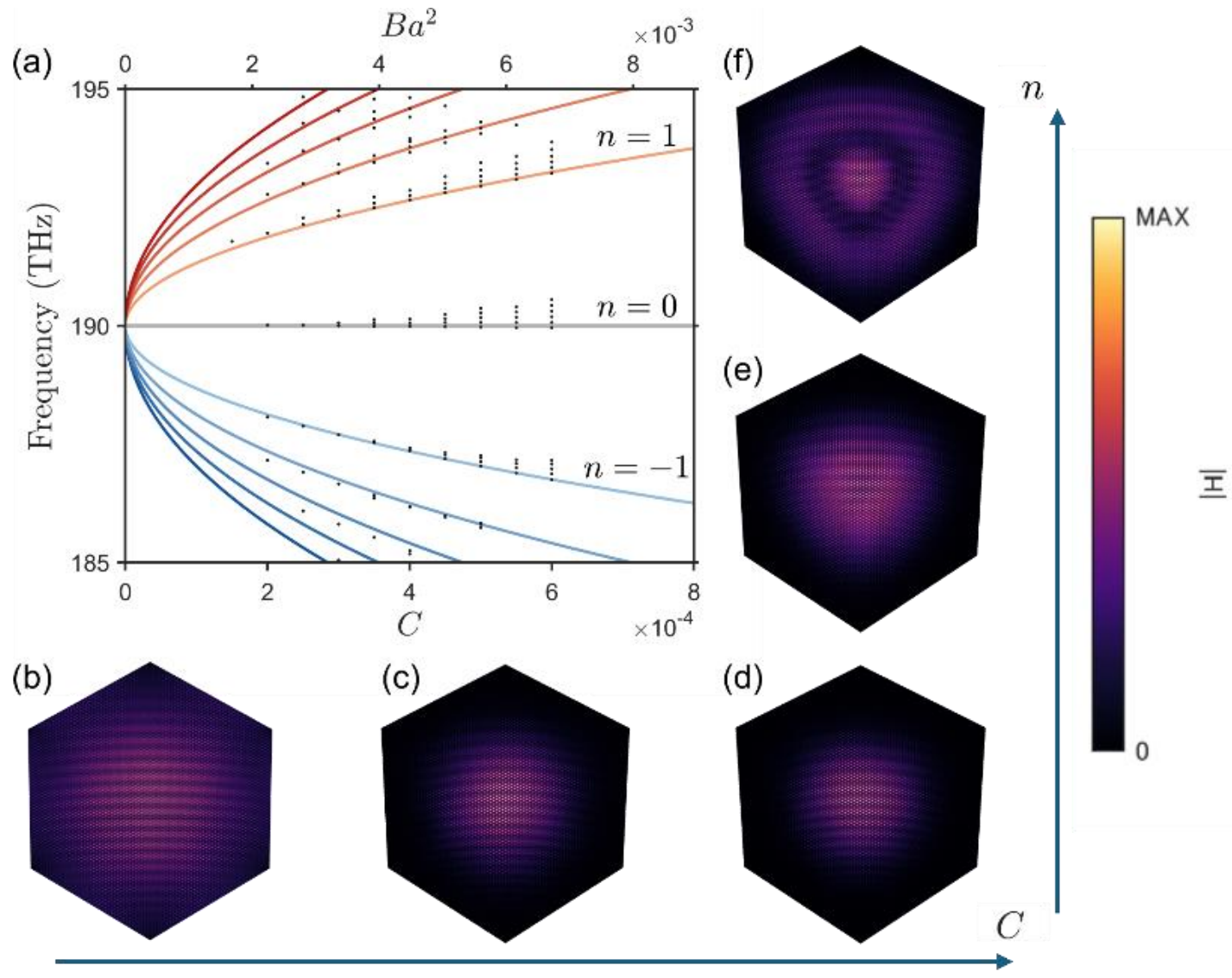

**Figure 3. Formation of photonic Landau levels in the triaxially strained photonic crystal.** (a) Simulated eigenfrequencies of the finite triaxially strained photonic crystal as a function of deformation strength C. Black dots denote the simulated eigenmodes, and solid curves indicate the theoretical prediction of Landau levels. (b-f) Representative field distributions of selected eigenmodes. Panels (b–d) show zeroth Landau level modes for $C = 2 \times 10^{-4}$, $4 \times 10^{-4}$, and $5 \times 10^{-4}$, respectively. Panels (e) and (f) show representative

higher-order Landau level modes with $n = 1$ and $n = 2$ at $C = 5 \times 10^{-4}$.

We then perform eigenfrequency simulations for finite hexagonal structures with different strain strengths $C$, corresponding to different pseudomagnetic fields $B_{\text{ps}} \propto C$. The results are shown in Figure 3a. As $C$ increases, the eigenfrequencies near the Dirac frequency reorganize into a set of well separated spectral branches. The separation between these branches increases with strain strength, consistent with the prediction from theory. By fitting the simulated eigenfrequencies, we identify several pseudo Landau level branches within the frequency window considered. The agreement confirms that the triaxial deformation produces Landau quantization in the graphene-like photonic crystal. It is worth mentioning that higher-order levels gradually deviate from the ideal Dirac prediction because they move out from the Dirac cone region. At the same time, each Landau level branch contains a set of split modes, indicating the presence of a fine structure beyond the ideal degenerate Landau level picture. Here we show the representing modes within each Landau level, shown in Figure 3b-f. As strain strength $C$ increases, the mode size decreases, as a result of decreasing of magnetic length. Also, the mode distribution also strongly depends on the index of Landau level, whose details are discussed in the Supplementary Information S2.

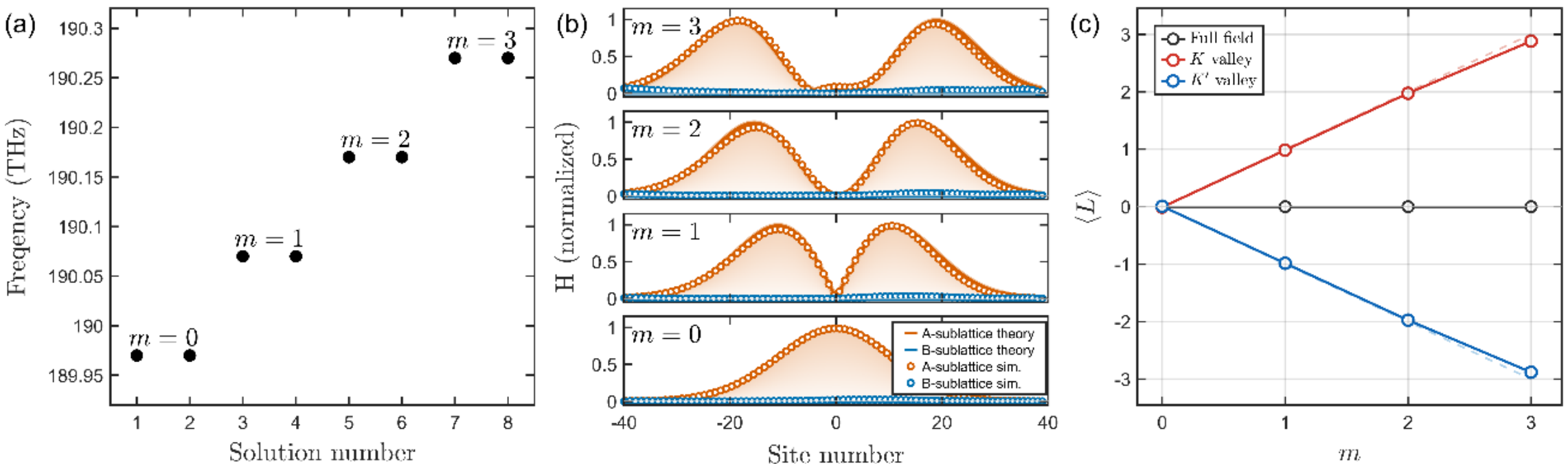

**Figure 4. Fine structure of the zeroth Landau level and valley-resolved angular momentum.** (a) Eigenfrequency of the split modes in zeroth Landau level. (b) Field intensity distributions of the corresponding split modes. Two sublattices are shown in different colors. (c) Valley-resolved angular momentum of the split modes. The K and K' projected components carry angular momenta of opposite sign, shown in red and blue, respectively, while the total angular momentum remains zero, shown in black.

Having established the formation of photonic Landau levels, we next examine the fine structure of the zeroth Landau level. In the ideal Dirac–Landau problem without a scalar potential, the zeroth Landau level is degenerate at the Dirac frequency. In the spectrum of the triaxial strained photonic crystal, however, several spectrally resolved modes emerge in the vicinity of this reference frequency. These sublevels form an approximately equally spaced frequency ladder, as shown in Figure 4a. This behavior is consistent with the quadratic scalar potential, which lifts the degeneracy of the Landau levels. Additionally, each sublevel appears as a nearly degenerate doublet. This double degeneracy originates from the two valleys, which experience pseudomagnetic fields of opposite signs. Since the scalar potential is valley independent, it preserves the valley paired degeneracy.

To identify the physical origin of these modes, we compare the radial field distributions obtained from eigenfrequency simulations with the envelope functions calculated from the effective Hamiltonian. For the deformation orientation and pseudomagnetic field sign used in the present structure, the zeroth Landau level modes are polarized on the A sublattice and can be written as

$$\Psi_{0,m} = \psi_{0,m}|A\rangle$$
$$\psi_{0,m} = \mathcal{N}_{0,m} e^{jm\phi} \left(\frac{r}{\sqrt{2}l_B}\right)^m e^{-\frac{r^2}{4l_B^2}} \tag{10}$$

where $\mathcal{N}_{n,m}$ is the normalized factor (see Supplementary Information S2). As shown in Figure 4b, the simulations reproduce this radial distribution, demonstrating quantitative consistency with the effective Hamiltonian description. In simulation results, all derived modes are strongly localized on the A sublattice, while their modal weight on the B sublattice is strongly suppressed. This sublattice polarization is a characteristic signature of the zeroth Landau level.[25] Moreover, position with maximum field intensity shifts with the resonance frequency. The lowest-frequency mode exhibits a central intensity maximum, whereas higher-index modes develop annular intensity distributions whose peak radii move progressively outward from the device center. This monotonic frequency–radius mapping constitutes a radial Landau rainbow. The corresponding results for higher-order Landau levels are provided in the Supplementary Information S4.

While the radial intensity distributions reveal the spatial ordering of the split modes, their angular structure requires a valley-resolved analysis. Unlike a Landau problem generated by a real magnetic field, the strain induced pseudomagnetic field reverses its sign between the two valleys. Consequently, the two valley-projected components of a given mode carry opposite angular momenta. In calculation, we first extract the envelope fields associated with the two valleys. Specifically, the complex magnetic field of the resonance state is Fourier filtered around the $K$ and $K'$ valleys, and the corresponding Bloch phases are removed to obtain the magnetic field envelop $H_\tau$, where $\tau$ denotes the $K$ or $K'$ valleys. We then define the valley-resolved angular momentum as

$$\langle L_\tau \rangle = \iint H_\tau^* \left(-j\frac{\partial}{\partial\phi}\right) H_\tau r dr d\phi \tag{11}$$

As shown in Figure 4c, the valley-resolved angular momenta provide direct evidence for the internal structure of the split Landau level. The two valleys contribute opposite angular momenta, reflecting the opposite signs of the strain induced pseudomagnetic field in the two valleys. Their magnitudes scale approximately as $m$, confirming that the split modes are organized by a valley-projected angular index. At the same time, the total angular momentum remains zero, indicating that the full electromagnetic eigenmodes are composed of balanced valley components. Thus, the pseudoelectric field resolves different angular sectors within the Landau level, while time-reversal-symmetry gives vanishing net orbital angular momentum in the observable mode.

**Discussion**

The intrinsic radial Landau rainbow demonstrated here provides a route to localized photonic states near the Dirac frequency, similar to existing Dirac vortex cavities.[26,27] However, unlike many Dirac vortex cavity designs that aim to realize an isolated single resonance, our structure produces a structured ladder of Landau modes. These resonances remain organized around the Dirac frequency while being spectrally separated.

Another closely related system is photonic Dirac cavities with spatially varying mass terms.[28] In the present case, the confinement does not originate from a mass induced potential well, but from the unfolding of Landau level degeneracy by the accompanying pseudoelectric field. As a result, the modes retain the characteristic sublattice polarization of the zeroth Landau level and remain tied to the Dirac frequency region, a feature that is not observed in mass confined Dirac cavities.

From an application perspective, the radial Landau rainbow offers a compact platform for spectrally addressable photonic resonances. Because the resonance frequency is correlated with the radial position of the modal intensity maximum, different modes could in principle be selectively coupled by placing a waveguide, emitter, gain region, or optical pump at different radial positions. This feature may be useful for spatially selective light–matter coupling[29–31], and mode-selective lasers[25]. Compared with lateral Landau-rainbow systems[20,21], where different frequencies are separated along an imposed gradient direction, the present structure realizes a radial frequency–position mapping without requiring an additional tuning.

**Conclusion**

We study the fine structure of a photonic Landau level in triaxial strained photonic crystals. The strain induced pseudomagnetic field first reorganizes the Dirac spectrum into Landau levels, while the accompanying pseudoelectric field lifts the degeneracy within the zeroth Landau level and resolves it into a sequence of discrete resonant modes. The nearly equal frequency spacing reflects the effective quadratic scalar potential, whereas the outward shift of the modal intensity maxima shows that different split modes are located in different radial regions of this potential. At the same time, the strong sublattice polarization confirms their zeroth Landau level origin. Valley-resolved angular momentum analysis further reveals that the two valleys carry opposite angular components, leading to vanishing net angular momentum in total.

Taken together, these results establish an intrinsic radial Landau rainbow in a strained photonic crystal. Unlike previously demonstrated lateral Landau rainbows, where frequency–position mapping is introduced by an imposed directional gradient, the present system generates a central organized radial frequency–position mapping through the accompanying pseudoelectric field in the triaxially strained photonic crystal. This mechanism converts Landau level degeneracy into a spectrally ordered and spatially addressable sequence of localized modes, providing a new route for designing Landau level based photonic resonances, multi-frequency cavities, and mode-selective light–matter interaction platforms.